
\magnification=\magstep1
\rightline{NO-ADRB/95/1}
\smallskip
\centerline{\bf A NOTE ON THE NEUTRINO DECAY LINE}

  \centerline{\bf AND THE POSSIBILITIES   OF ITS DETECTION} \smallskip

\centerline{\bf Srdjan Samurovi\' c$^1$ and Vladan \v Celebonovi\' c$^2$}
\medskip \centerline{\it $^1$ People's Observatory, Gornji grad 16, }

\centerline{\it Kalemegdan, 11000
Belgrade, Yugoslavia}
\centerline{\it e-mail: srdjanss@mrsys1.mr-net.co.yu}
\smallskip
\centerline{\it $^2$ Institute of Physics, Pregrevica 118, }

\centerline{\it  11080 Zemun--Belgrade, Yugoslavia}

\centerline{\it e-mail: vladanc@mrsys1.mr-net.co.yu}

\bigskip

When talking about spectral line shapes and intensities, one usually envisages
some plasma similar to those produced in a laboratory, or existing in a
natural environement. It is more than often completely disregarded that
spectral lines can occur in a seemingly non-plasmatic part of physics --
elementary particle physics.

The purpose of this contribution is to propose and discuss some mathematically
simple but physically important results  concerning  the  neutrino line and
the possibilities of its detection. The cosmological  importance  of neutrinos
stems from the fact that massive neutrinos may be the solution to the dark
matter  problem in the Universe (discussed in Frampton and Vogel, 1982;
Sciama, 1993; Gelmini and Roulet, 1994,  and numerous other publications).
Outside cosmology, neutrinos are useful as probes of solar and stellar
internal structure (for example, Hirata {\it  et al.\/}, 1987; Burrows, 1990;
Bahcall, 1994; Petcov, 1995) and stellar evolution (Castellani and
Degl'Innocenti, 1993). Various kinds of interactions in which neutrinos
participate can give rise to different spectral lines. The predicted
wavelengths depend on the supposed neutrino masses.

The modern particle physics theory allows that neutrinos could have rest
masses different from zero (Zeldovich and Hlopov, 1981). There are three
neutrino types: $\nu _e$, $\nu _\mu$ and $\nu _\tau$, i.e. the number of types
is equal to: $N_\nu =3.00\pm 0.04$ (Ting, 1993). The upper limit for their
masses based upon earth based experiments are:
$$m_{\nu _e}<o(10\, {\rm eV}),$$
$$m_{\nu _\mu}<160\, {\rm KeV}$$
and
$$m_{\nu _\tau}<31\, {\rm MeV}.$$

\noindent (Gelmini and Roulet, 1994).

The hot big-bang cosmology sets the upper limit to the total neutrino mass
(Klink\-hamer and Norman, 1981):
$$\sum _i m_{\nu _i}< 100\, {\rm eV}.$$

Neutrino mass plays the central role in the Scia\-ma's (1993)
decaying dark matter (DDM) model as well as in cold + hot dark
matter  (C$\nu ^2$DM) model created by Primack, Holtzman, Klypin
and Caldwell (1995).

Neutrinos  were relativistic at the decoupling  and  their present
density for each neutrino type is: $$n_\nu =112\, {\rm cm}^{-3}$$ (Sciama,
1993).

The value of the neutrino mass for the type $x$, where $x=\nu , \mu, \tau$ can
be expressed in the following form (see for example Schaeffer, 1994 and
Samurovi{\'c}, 1995): $$m_{\nu _x}\sim 94 h^2\, \Omega _{\nu _x} \, {\rm
eV}\eqno(1)$$ where $\Omega _{\nu _x}$ is the density parameter for the
universe, and the estimations for its value are different in the different
models. Hubble parameter, $h\equiv {H_0\over 100 {\rm km/s\over MPc}}$ varies
between 0.5 and 1, but it is possible that it is close to 0.5 in order
not to get too small a value of the age of the universe.

Thus DDM model predicts $\Omega _{\nu _x}\sim 1$ and

$$m_{\nu _x}\sim 30\, {\rm eV}$$ (Sciama, 1993). Here $\nu _x$
denotes most probably tau neutrino, $\nu _\tau$.
Sciama (1993) found a lower limit for $m_\nu$ for the Milky Way:
$$m_\nu=27.6\pm 1\, {\rm eV}.$$

Because of the fact that different neutrino types have different rest masses
it is possible that the heavier neutrino decays into a lighter one. This
process is followed by the emission of photons which produces observable
ionization effects. Neutrino decay can be represented in the
following form:
$$\nu _1\rightarrow \nu _2+\gamma.$$
We shall take that the more massive tau neutrino, $\nu _\tau$, decays into a
photon and a  muonic neutrino, $\nu _\mu$, where $m_{\nu _\tau}\gg m_{\nu
_\mu}$: $$\nu _\tau\rightarrow \nu _\mu +\gamma, \> \, \, \, E_\gamma={1\over
2}E_{\nu _\tau}={1\over 2}m_{\nu _\tau}.\eqno(2)$$

DDM theory (Sciama, 1993) gives the folowing value of this ratio ${m_{\nu
_\mu}\over m_{\nu _\tau}}\ ^<_\sim 0.2$. If we take that $E_{\nu
_\tau}=28\, {\rm eV}$ we shall get the flux of 14 eV photons -- this value
exceeds a little the Lyman limit of 13.6 eV for the hydrogen ionizing
energy. However, these emitted photons will be redshifted by cosmological
expansion and, therefore, will be able to ionize hydrogen near their point of
origin. Also, self-absorption of the decay photons by the neutrinos can be
ignored, because their energy will be below the one needed to initiate ``the
reverse action''of the type $\nu _\mu +\gamma\rightarrow \nu _\tau$ (OWB).

Due to the similarity of their ionization potentials,
 wherever hydrogen is ionized, nitrogen is also ionized.
It can be shown (Sciama, 1993a) that:
 $$14.53<E_\gamma < 14.68\, {\rm eV}.$$

The upper limit for the value of the decaying photon energy follows from the
requirement that  these photons are the solution of the ${\rm
C^0/CO}$ ratio problem.

Thus one obtains  that
$$E_\gamma=14.605\pm 0.075\, {\rm eV}$$
(Sciama, 1993a). This would give, according to  equation (2) the
following value for the tau neutrino mass: $$m_{\nu _\tau}=29.21\pm 0.15\, {\rm
eV}$$ (Sciama, 1993a).

Under the assumption that the dark halo of the Galaxy consists mainly of
massive neutrinos, Melott and Sciama (1981) and Sciama (1993) found that the
neutrino decay will give the following limit for the neutrino lifetime:
$$\tau\ge 10^{23} {T \overwithdelims () 10^4}^{3\over 2} {0.6\, {\rm cm}^{-3}
\overwithdelims () n_e}^2 {1\, {\rm kpc} \overwithdelims () d}\times$$
$${0.05\, {\rm  rad}\overwithdelims () \varphi } {30\, {\rm eV}\overwithdelims
() m_\nu} N$$ where $\tau$  is expressed in seconds. $T$ is the cloud
temperature, $n_e$ its electron density, $d$ its distance, $\varphi$ its
angular radius and $N$ is the average number of ionizations caused by a given
photon.  From this equation one can see that lifetime $\tau$  ranges from
$10^{23}$ to $10^{24}$ s.

One can establish an equation that connects the lifetime $\tau$ and
transition magnetic moment $\mu _{12}$ under the assumption
${m_{\nu _\mu}\over m_{\nu _\tau}}\ll 1$ (Sciama, 1993):
$$ \tau = 10^{23} {30\, {\rm eV} \overwithdelims () m_1}^3 {10^{-14} \mu _B
\overwithdelims () \mu_{12}}^2 {\rm s}$$
where the Bohr magneton $\mu _B={eh\over 2m_e c}$. For $m_1=m_{\nu_\tau}\sim
30$ eV and $\mu_{12}\sim 10^{-14} \mu_B$ one would obtain the lifetime $\tau
\sim 10^{23}$ s. Castellani and Degl'Inno\-centi (1993) considered  the
effect of a nonvanishing neutrino magnetic moment on the stellar evolution and
obtained the upper limit $\mu_{12}<10^{-12}\mu_B$ which points out that
the estimated value for the neutrino lifetime $\tau$ is possible.

Although Hopkins Ultraviolet Telescope during its observations of
the cluster A655 did not detect the line of the decay photon with
energy $E_\gamma$ that ranges between 14.5 and 15~eV (Sciama,
1993) it is quite possible that the dark matter in the center of
this cluster is partly baryonic (Ashman, 1992). In this case the
strength of the neutrino decay  line is less than the one
predicted by the fact that {\it all\/} the dark matter in the
cluster is made of decaying neutrinos (Ashman, 1992).  In this
cluster there could be significant amounts of neutrinos (Sciama,
1993).

However, Reimers and Vogel (1993) detected HeI resonance lines in four high-
redshift Lyman limit systems of the QSO HS 1700+6416 (z=2.72). They found
neutral hydrogen to neutral helium column density ratios ${N_{\rm HI}\over
N_{\rm HeI}}\approx 30$. According to their analysis HeI column densities are
$\sim 5$ times greater than column densities obtained by their model
calculations. The fact that effective hydrogen ionizing flux is
$\sim 8$ times greater than helium ionizing flux may be the consequence of the
feature of the DDM theory, according to which decay photons can ionize
hydrogen but not helium (Sciama, 1994).

According to C$\nu ^2$DM theory proposed by Primack, Holtzman,
Klypin and Caldwell (1995) that uses cold + hot dark matter
(CHDM) requirement that a total neutrino mass $\sim$ 5 eV,
suggests that masses of mu and tau neutrinos are approximately
equal i.e. $m_{\nu _\mu}\approx m_{\nu _\tau} \approx 2.4$ eV. It
is worth noting that this value is obtained from equation (1),
but the density parameter is taken to be $\Omega _\nu\sim 0.3$
and the Hubble parameter is again $h\sim 0.5$.

We discuss the simple case of the radiative decay of these light
neutrinos. The obvious relation:
$$m_\nu c^2=h{c\over \lambda}$$ gives $$\lambda = {h\over m_\nu c}$$
for the wavelength of the spectral line emitted in the process.
When we insert the appropriate numerical values, we obtain:
$$\lambda_{\rm nm}={1239.85\over m_{\nu [{\rm eV}]}}.$$
For $m_\nu=2.4$ eV, we obtain $\lambda=516.6$ nm.

Another spectroscopically interesting question is the shape of
this line. As cross sections for neutrino interactions with
matter are very low, it can safely be assumed that the line is
very nearly mono\-chro\-matic, i.e., that it only has the natural
width. Its value can be simply estimated from the uncertainty
principle as:
$$\Delta \lambda ={\lambda ^2 \over 2\pi c \tau}$$
which tends to 0 due to huge values of $\tau$.

The order of magnitude estimates of the neutrino decay line
wavelength  are presented in the
Table 1.

\noindent {\bf Table 1}
\smallskip

\vbox{\tabskip=0pt \offinterlineskip
\def\podvuci{\noalign{\hrule}}
\def\razmak{\noalign{\vskip.1cm}}
\halign to \hsize{\strut#& \vrule#\tabskip=0pt  plus 10pt minus5pt &
\hfil#\hfil&\vrule#& \hfil#&\vrule#& \hfil#&\vrule#
\tabskip=0pt\cr\podvuci
&& {\rm Neutrino type}&& $m_\nu$&&  $\lambda$ {\rm [nm]}
&\cr\podvuci\razmak\podvuci

 &&$e$   &&   10 \ \ eV&&123.985& \cr
 && $\mu$ && 150
KeV&&0.0083& \cr
&& $\tau$&& 30 MeV&&$4.1\times 10^{-5}$&\cr
\podvuci}}

\medskip

It can be concluded that the line is monochromatic. As for the wavelenghts,
the only one which could be experimentally detected is the line originating in
the radiative decay of $\nu _e$. The precise value of the wavelength depends
(obviously) on the mass; if $m_\nu \approx 2.4$ eV, as in the C$\nu ^2$DM
theory, $\lambda=516.6$ nm. The line intensity is a function of the
concentration of the decaying neutrinos; the precise form of this dependence
remains a problem for further study. It has very recently been shown
(Vassilevskaya, Gvozdev and Mikheev, 1995) that the probability of a radiative
decay of a massive neutrino can greatly increase in strong electromagnetic
fields. Under certain conditions, the increase can be as large as $10^{33}$
times! This renders the detection of the neutrino decay line much easier in
the case of astronomical objects in whose vicinity exist strong
electromagnetic fields.

A word about cosmological neutrinos. A neutrino decaying in an
astronomical object having redshift $z$ would give a line with

$$\lambda={h\over m_\nu c}(1+z)$$
which could be important in case of QSOs, for example.

\bigskip

{\bf  CONCLUSION}

\bigskip
The object of this note is the spectral line due to the radiative
decay of a massive neutrino. We have calculated the wavelength
and the natural width of the line as a function of the mass and
lifetime of the neutrino having $m_\nu
\ ^<_\sim 10$
eV. Several observational consequences of this process were
discussed. The newly discovered enhacement of the decay rate in
strong electromagnetic fields renders the analysis of the
neutrino decay line important even for observational astronomers.
\bigskip

{\bf REFERENCES}
\bigskip

\item{}\kern-\parindent{Ashman, K. M. 1992, {\it Pub A.S.P.\/},
{\bf 104}, 1109.}

\item{}\kern-\parindent{Bahcall, J. N.: 1994, {\it Phys.
Lett.\/}; {\bf B338}, 276.}

\item{}\kern-\parindent{Burrows, A.: 1990, {\it Annu. Rev. Nucl.
Part. Sci.}, {\bf 40}, 181.}

\item{}\kern-\parindent{Hirata, K., Kajita, T., Koshiba, M. {\it et al.}: 1987,
{\it Phys. Rev.
Lett.}, {\bf 58}, 11490.}

\item{}\kern-\parindent{Castellani, V. and Degl'Innocenti, S.:
1993, {\it Ap. J.}, {\bf 402}, 574.}

\item{}\kern-\parindent{Frampton, P. H. and Vogel, P.: 1982, {\it
Phys. Rep.}, {\bf 82}, 339.}

\item{}\kern-\parindent{Gelmini, G. and Roulet, E.: 1994,
preprint  CERN-TH 7541/94.}

\item{}\kern-\parindent{Melott, A. L.  and Sciama, D. W.: 1981,
{\it Phys. Rev. Lett.}, {\bf 46}, 1369.}

\item{}\kern-\parindent{Overduin, J. M., Wesson, P. S. and
Bowyer, S. (OWB): 1993, {\it
Ap. J.}, {\bf 404}, 460. }

\item{}\kern-\parindent{Petcov, S. T.: 1995, preprint SISSA
12/95/EP.}

\item{}\kern-\parindent{Primack, J. R., Holtzman, J., Klypin, A. and Caldwell,
D. O.: 1995, {\it Phys. Rev. Lett.}, {\bf 74}, 2160.}

\item{}\kern-\parindent{Reimers, D. and Vogel, S. 1993, {\it A\&
A}, {\bf 276}, L13.}

\item{}\kern-\parindent{Samurovi{\'c}, S.: 1995, unpublished
work.}

\item{}\kern-\parindent{Schaeffer, R.: 1994, {\it Cosmologie\/},
Saclay  preprint, T94/025.}

\item{}\kern-\parindent{Sciama, D. W.: 1993, {\it Modern Cosmology and the
Dark Matter Problem}, Cambridge University  Press.}

\item{}\kern-\parindent{Sciama, D. W.: 1993a, {\it Ap. J.}, {\bf
415}, L31.}

\item{}\kern-\parindent{Sciama, D. W.: 1994, {\it Ap. J.}, {\bf
422}, L49.}

\item{}\kern-\parindent{Ting, S. C. C.: 1993, preprint CERN -
PPE/93-34.}

\item{}\kern-\parindent{Vassilevskaya, L. A., Gvozdev, A. A. and
Mikheev, N. V.: 1995, {\it Jadernaja fizika\/}, {\bf 58}, 712.}

\item{}\kern-\parindent{Zeldovich, Y. B. and Hlopov, M. Y.: 1981,
 {\it U.F.N.}, {\bf 135}, 45.}

\bye